%% file: main.tex
\documentclass[10pt,twocolumn]{article}

\usepackage[margin=1in]{geometry}

\usepackage{amsmath,amssymb,amsfonts}
\usepackage{algorithm}
\usepackage{algorithmic}
\usepackage{multirow}
\usepackage{booktabs}
\usepackage{xcolor}

\usepackage{colortbl}
\usepackage{hhline}
\usepackage{array}
\usepackage{makecell}
\usepackage{tabularx}
\usepackage{adjustbox}

\usepackage{graphicx}
\usepackage{wrapfig}
\usepackage{tikz}
\usepackage{pgfplots}
\pgfplotsset{compat=1.17}
\usepackage{stfloats}
\usepackage{tcolorbox}
\usepackage{pifont}
\usepackage{enumitem}

\usepackage[numbers,sort&compress]{natbib}

\usepackage[colorlinks=true,citecolor=blue,linkcolor=blue,urlcolor=blue]{hyperref}

\newcommand{\ours}{\textsc{HeadRouter}}

\newcommand{\sel}{\mathrm{sel}}
\newcommand{\spr}{\sigma_{\sel}}

\definecolor{mygray}{gray}{.92}
\definecolor{ForestGreen}{RGB}{34,139,34}
\definecolor{Forestred}{RGB}{220,50,50}
\definecolor{darkpink}{RGB}{255, 20, 147}

\newcommand{\fg}[1]{\mathbf{\textcolor{ForestGreen}{#1}}}

\definecolor{delectricblue}{RGB}{93, 117, 131}
\definecolor{tkcolor2_back}{RGB}{249,237,238}
\definecolor{tkcolor2_frame}{RGB}{244,136,146}
\definecolor{oursblue}{RGB}{237,247,252}
\colorlet{lightdelectricblue}{delectricblue!30}

\providecommand{\Description}[1]{}

\begin{document}

\title{\ours{}: Dynamic Head-Weight Routing for Task-Adaptive Audio Token Pruning in Large Audio Language Models}

\author{
  Peize He$^{1,2}$, Yaodi Luo$^{1,2}$, Xiaoqian Liu$^{1,3}$, Xuyang Liu$^{1,4}$, Jiahang Deng$^{1}$, \\
  Yaosong Du$^{2}$, Li Bangyu$^{2}$, Xiyan Gui$^{1,5}$, Yuxuan Chen$^{1}$, Linfeng Zhang$^{1}$\thanks{Corresponding author.} \\[1.4ex]
  \normalsize $^{1}$EPIC Lab, Shanghai Jiao Tong University \quad
  $^{2}$DAIL Tech \quad
  $^{3}$Northeastern University \\[0.2ex]
  \normalsize $^{4}$Sichuan University \quad
  $^{5}$Huazhong University of Science and Technology \\[0.2ex]
  \normalsize \textbf{Homepage:} 
  \href{https://dabdans.github.io/HeadRouter/}{\texttt{https://dabdans.github.io/HeadRouter/}}
}

\date{}

\maketitle

\begin{abstract}
Recent large audio language models (LALMs) demonstrate remarkable capabilities in processing extended multi-modal sequences yet incurs high inference cost.
Token compression is an effective method by directly reducing redundant tokens in the sequence.
Existing compression methods usually assume that all attention heads in LALMs contribute equally to various audio tasks, calculate token importance by averaging scores across all heads. However, our analysis demonstrates that attention heads exhibit distinct behaviors across diverse audio domains.
We further reveal that only a sparse subset of attention heads actively responds to audio, with completely different performance when handling semantic and acoustic tasks.
In light of this observation, we propose \ours{}, a head importance aware token pruning method that perceives the varying importance of attention heads in different audio tasks to maximize the retention of crucial tokens.
\ours{} is training free and can be applied to various LALMs.
Extensive experiments on the AudioMarathon and MMAU-Pro benchmark demonstrate that \ours{} achieves state-of-the-art compression performance, exceeding the baseline model even when retaining \textit{70\%} of the audio tokens, achieving \textbf{101.8\%} and \textbf{103.0\%} of the vanilla average on the Qwen2.5-Omni-3B and Qwen2.5-Omni-7B, respectively.
\end{abstract}

\textbf{Keywords:} Audio LLM, Multi-modal Large Language Models, Efficiency, Long-context

\input{section/1_introduction}

\input{section/2_related_work}

\input{section/4_method}

\input{section/5_experiments}

\input{section/6_discussion}
\input{section/7_conclusion}

\bibliographystyle{plainnat}
\bibliography{headrouter}

\end{document}

%% file: section/1_introduction.tex
\definecolor{mygreen}{RGB}{152,190,118}
\definecolor{caption_orange}{RGB}{214,169,109}
\definecolor{caption_green}{RGB}{159,200,148}

\section{Introduction}
\label{sec:intro}
Large Audio-Language Models (LALMs)~\citep{cui2025recent,tang2023salmonn,xu2025qwen3,goel2025audio}
have recently emerged as a powerful paradigm for general-audio understanding.
By directly coupling an audio encoder with a pretrained Large Language Model (LLM) via a lightweight projection module, LALMs achieve joint modeling of complex acoustic cues and deep linguistic meaning~\citep{abouelenin2025phi,xu2025qwen3,goel2025audio}. This end-to-end design bypasses the bottlenecks associated with traditional, rigid cascade systems such as ASR (Automatic Speech Recognition)-to-text pipelines~\citep{radford2023robust,zhang2023speechgpt}, enabling richer interaction between modalities. However, this advancement comes at a steep computational cost: processing long audio inputs generates massive token sequences, leading to prohibitive prefilling latency and excessive KV-cache memory footprints during inference. This severe efficiency bottleneck sharply limits the practical deployment of LALMs for long-context applications~\citep{xiao2023efficient,liu2025mixing,ghosh2025audio,liu2025shifting}.

\begin{figure}[t]
  \centering
  \includegraphics[width=\linewidth]{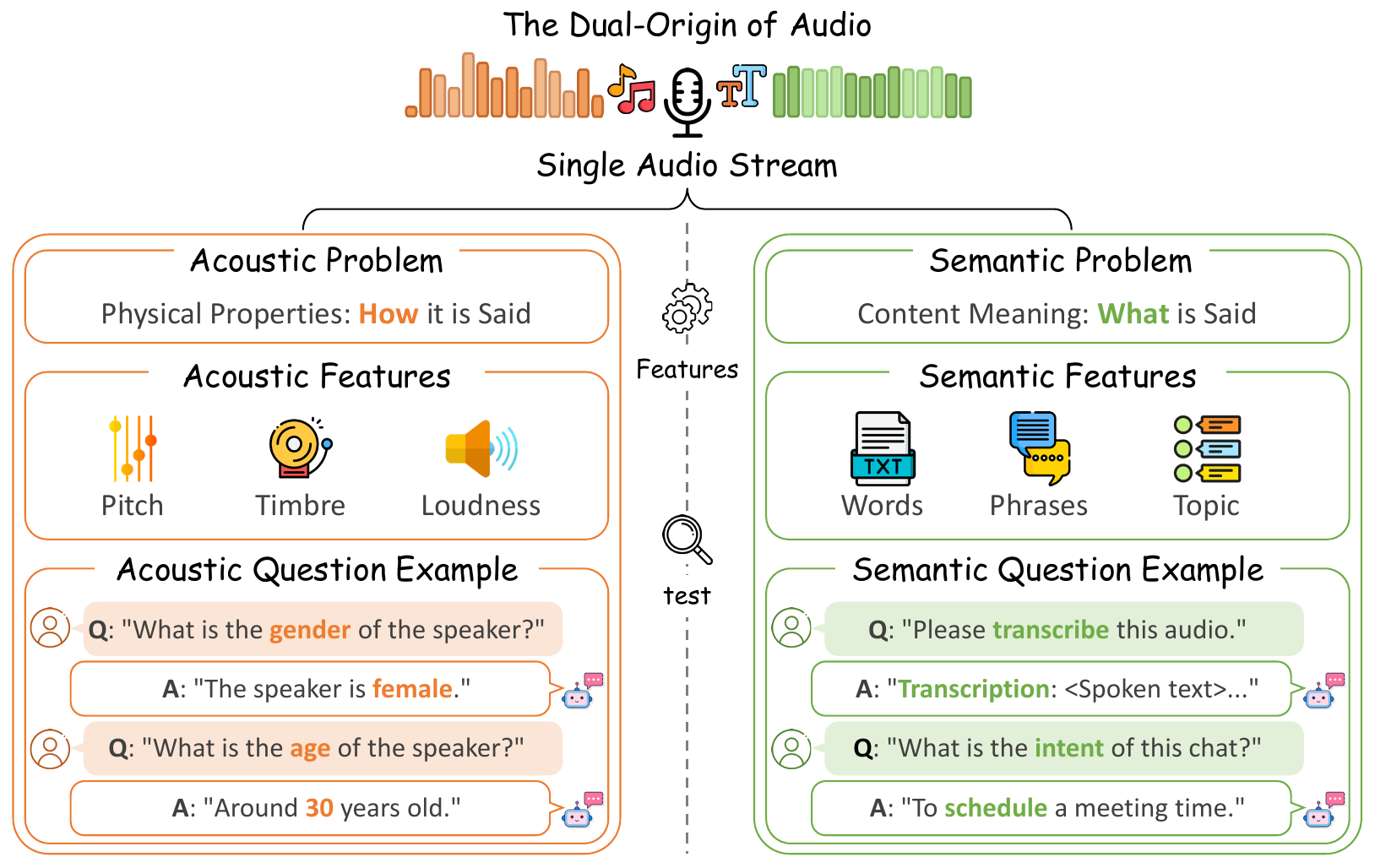}
  \caption{Overview of the audio task categories evaluated in this work. \textcolor{orange}{Left:} acoustic tasks. \textcolor{mygreen}{Right:} semantic tasks.}
  \vspace{-4mm}
  \label{fig:task-overview}
  \Description{Overview schematic of the heterogeneous audio tasks evaluated.}
\end{figure}

To mitigate this problem, recent studies have introduced token compression methods that directly reduce the number of audio redundant tokens~\citep{bhati2025towards,he2025audiomarathon,wang2026audiokv,ding2026omnisift}. Existing research on audio token compression can be mainly classified into three principal categories: similarity-based, temporal-based, and attention-based approaches~\citep{chen2024image,bhati2025towards}. Similarity-based~\citep{wen-etal-2025-stop,he2025audiomarathon} techniques identify and prune redundant token regions by computing similarity across audio embeddings. However, these approaches tend to mistakenly discard similar yet meaningful information within the audio, causing misunderstandings. Temporal-based subsampling methods~\citep{he2025audiomarathon}, simply operating in a context-agnostic manner. Meanwhile, they retain an excessive amount of background noise, while entirely disregarding user instructions. Attention-based methods~\citep{bhati2025towards,chen2024image}, in turn, derive token importance scores from attention weights. These approaches typically presuppose equal contribution from all attention heads to various audio tasks and simply use an arithmetic average weight over all heads, overlooking the differences in the capabilities of each attention head in audio scene.

In terms of audio scene, our analysis identifies \textbf{\textit{three fundamental issues}} overlooked by existing pruning methods: \textbf{\textit{(1) distinct task heterogeneity}}. Audio scene contains a diverse range of tasks, as illustrated in Figure~\ref{fig:task-overview}, ranging from semantic understanding~\citep{wang2025mmsu,yang2025sakura,kumar2025mmau}, including automatic speech recognition~\citep{panayotov2015librispeech}, to acoustic analysis, including event detection~\citep{serizel2020sound} and speaker identification~\citep{chengvoxdialogue,kumar2025mmau}. \textbf{\textit{(2) head-behavior heterogeneity}}. When processing different audio signals, as shown in Figure~\ref{fig:tsne} and Figure~\ref{fig:selectivity-heatmap}, attention heads inherently exhibit distinct importance depending on the specific domain of the problem. During our investigation into these head behaviors, we discovered a profound difference in head-weight distributions between \textbf{\textit{semantic and acoustic tasks}}. \textbf{\textit{(3) lack of audio nature awareness}}. FastV~\citep{chen2024image} exhibits severe positional bias, retaining tokens at the tail while discarding earlier segments. Frame~\citep{he2025audiomarathon} subsampling is content-blind, unable to distinguish voiced speech from silence. From the pruning results of oracle, which yields a natural pattern of audio, revealing that pruning must be audio-aware, adapting to voice structure rather than positional heuristics or uniform sampling.

\begin{figure}[t]
  \centering
  \includegraphics[width=0.95\linewidth]{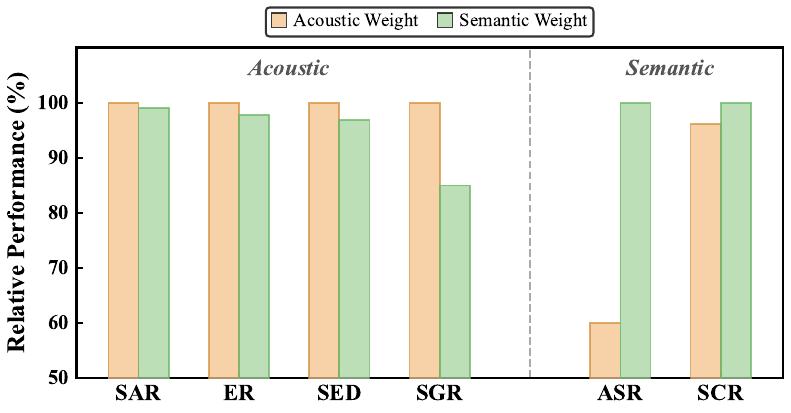}
  \vspace{-2mm}
  \caption{Comparisons on performance of semantic and acoustic tasks for two head weight profiles: \textcolor{caption_orange}{acoustic profile} is weight for token pruning on acoustic tasks, \textcolor{caption_green}{semantic profile} is weight for token pruning on semantic tasks, demonstrating different kinds of tasks require proper profiles.}
  \label{fig:compare-local}
  \vspace{-4mm}
  \Description{Bar chart showing performance advantage of semantic over acoustic head-weight profiles across tasks at two pruning ratios.}

\end{figure}

\begin{figure}[t]
  \centering
  \includegraphics[width=\linewidth]{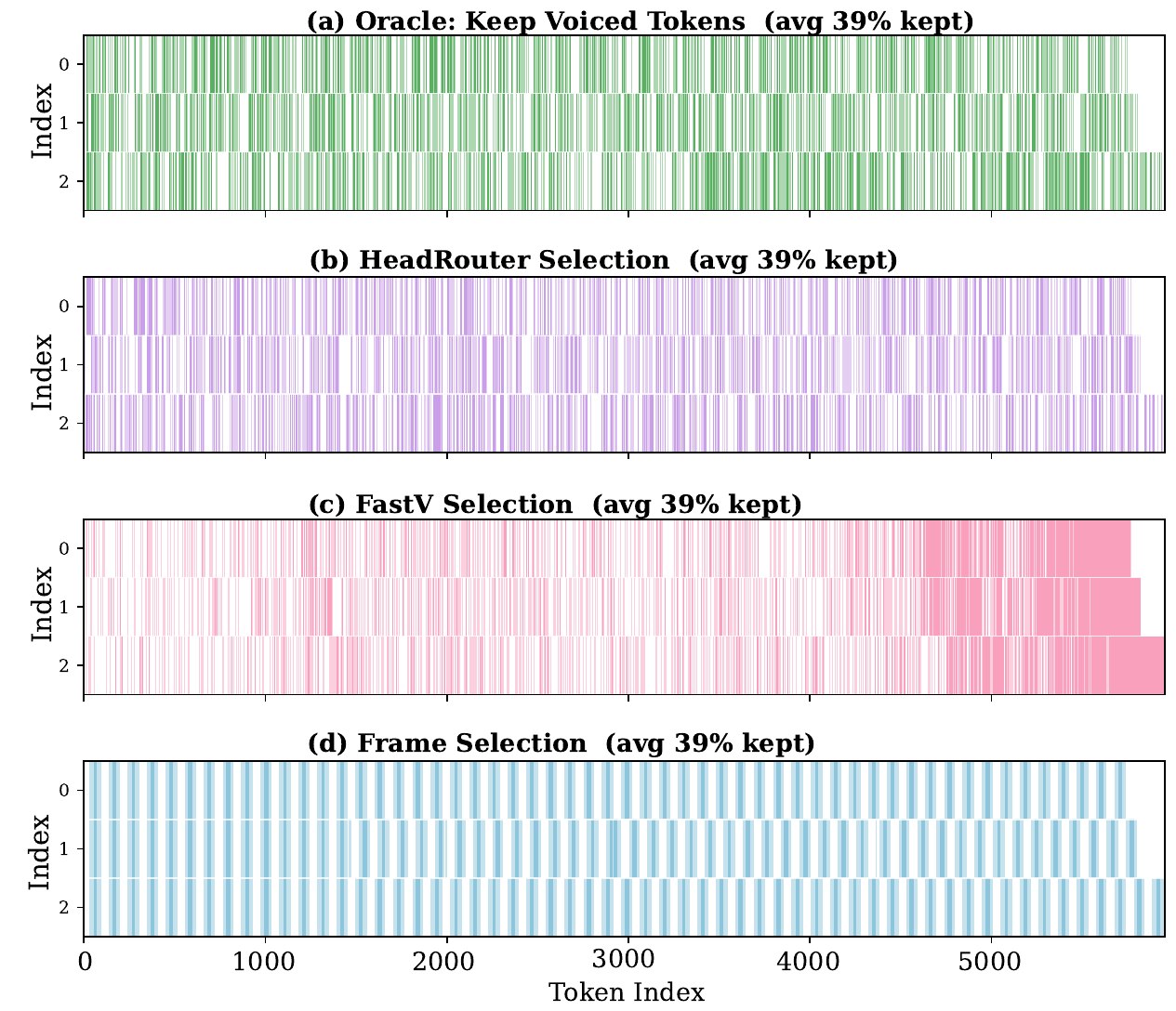}
  \vspace{-6mm}
  \caption{\textbf{Selected tokens} on 3 samples with different methods. The oracle result is obtained by token with higher energy.}
  \label{fig:oracle-r06}
  \Description{Bar chart comparing HeadRouter against oracle routing and other pruning methods at ratio 0.6 across AudioMarathon tasks.}
  \vspace{-4mm}
\end{figure}

Based on the issues analyzed above, we argue that audio tasks are inherently heterogeneous (issue 1), a one-size-fits-all pruning strategy is unsuitable to handle certain task families, as evidenced by Figure~\ref{fig:compare-local} on AudioMarathon~\citep{he2025audiomarathon}. Notably, \textbf{\textit{issue (2) offers a natural remedy}}: since attention heads already exhibit task-dependent behavior, their heterogeneous activation patterns can serve as an intrinsic signal for adaptive head-weight routing, enabling the pruning strategy to dynamically match the demands of each task. Furthermore, beyond task-adaptive head weighting, effective pruning must also be audio-aware (issue 3). As illustrated in Figure~\ref{fig:oracle-r06}, the oracle retains only voiced segments in a sparse, non-uniform pattern, fundamentally different from any positional or temporal heuristic. Taken together, these observations motivate a strategy that leverages \textit{\textbf{head-behavior heterogeneity}} for \textit{\textbf{task-adaptive}} routing, while simultaneously achieving \textit{\textbf{audio-aware}} token selection.

To address these challenges, we introduce \textbf{\ours{}}, a training-free, per-sample dynamic head-weight routing method for robust and adaptive audio token pruning. Our method operates by first computing an entropy-based \textit{selectivity} score for each head to measure how concentrated its attention is over audio tokens. The standard deviation of it across all heads then serves as a compact routing signal that captures whether the input is semantic-like {\textit{(low)}} or acoustic-like \textit{(high)}. Based on this signal, \ours{} performs Gaussian soft mixing over three pre-calibrated distinct head-weight profiles: semantic, uniform, and acoustic, producing a continuously head-weight profile for final token scoring. This continuous and smooth routing strategy preserves task-critical information.

We evaluate \ours{} across two massive, comprehensive benchmarks: AudioMarathon~\citep{he2025audiomarathon} and MMAU-Pro~\citep{kumar2025mmau}. Experiments demonstrate that \ours{} consistently matches or outperforms strong fixed-profile and other state-of-the-art pruning baselines, with constant performance gains observed under aggressive pruning ratios across benchmarks. These results highlight that input-adaptive head routing is a practical and effective principle.

Our contributions are summarized as follows:
\begin{itemize}[leftmargin=*, itemsep=0pt, parsep=0pt, topsep=0pt]
\item \textbf{Semantic–acoustic head divergence.} We provide an in-depth analysis of head-behavior heterogeneity across diverse audio tasks, characterizing a clear semantic-acoustic divergence.
\item \textbf{Audio-adaptive routing.} We introduce \ours{}, a novel training-free dynamic routing mechanism based on no-RoPE selectivity statistics and Gaussian soft profile mixing.
\item \textbf{Performance-enhancing compression.} Across multiple benchmarks and models, \ours{} are able to exceed the unpruned baseline when removing 30\% of audio tokens. For instance, \textbf{+1.2} points and \textbf{+2.1} points on the AudioMarathon~\citep{he2025audiomarathon} average for Qwen2.5-Omni-3B and Qwen2.5-Omni-7B respectively, while all competing methods suffer clear degradation.

\end{itemize}

\begin{figure*}[t]
  \centering
  \includegraphics[width=\textwidth]{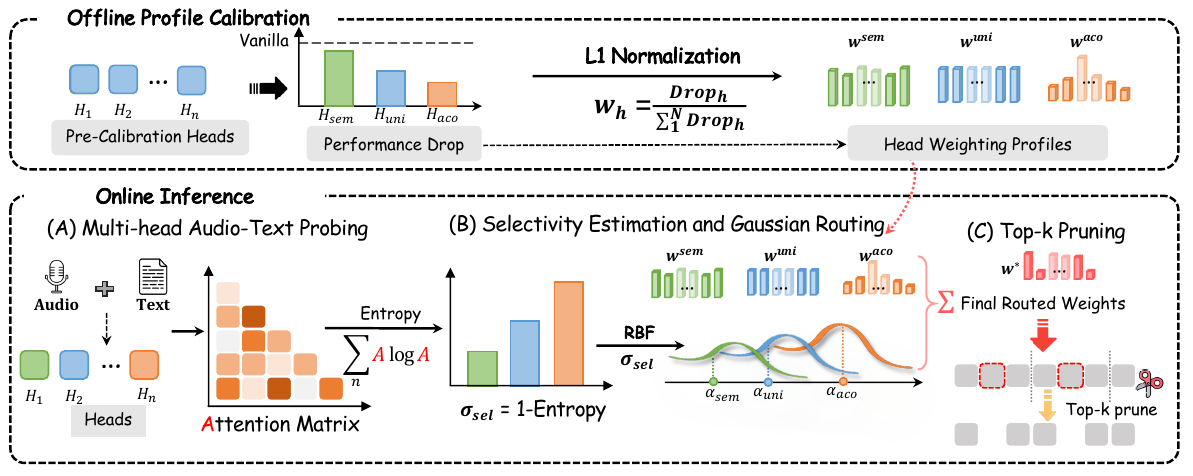}
 \caption{Overview of \ours{}. A QK probe at layer $M$ extracts the selectivity spread ($\spr$) from tokens processed by preceding layers. This spread is fed into a radial basis function (RBF) kernel, $\exp\!\bigl(-\lVert s - \mu_k \rVert^2 / 2\sigma^2\bigr)$
, where each $\mu_k$ anchors a profile center and the normalized kernel outputs serve as mixing weights.}
  \label{fig:overview}
  \Description{System diagram showing the \ours{} pipeline with
    Qwen2.5-Omni.}

\end{figure*}

%% file: section/2_related_work.tex
\section{Related Work}
\label{sec:related}

\noindent \textbf{Large Audio-Language Models.}
Large Audio-Language Models (LALMs) achieve multimodal understanding by integrating an audio encoder, a projection module, and a pretrained LLM~\citep{cui2025recent,he2025audiomarathon,qwen2.5-omni,ghosh2025audio}. Comprehensive benchmarks such as AudioMarathon~\citep{he2025audiomarathon} and MMAU-Pro~\citep{kumar2025mmau} have been proposed to evaluate LALMs across diverse audio understanding capabilities, spanning speech, sound, music, and spatial audio domains. In real-world scenarios, however, processing minute-scale audio  in realistic long context settings, such as meetings and podcasts, results in an immense influx of audio tokens: one minute of audio is embedded into 1{,}500 tokens with Whisper-based encoders~\citep{he2025audiomarathon,qwen2.5-omni,abouelenin2025phi}, due to simple frame-level downsampling or hard splitting of the audio signal. This proliferation of tokens drastically increases computational complexity, particularly during the \emph{prefilling} phase, and thus critically hinders the vision of seamless, low-latency voice interactions.

\noindent \textbf{Token Pruning: From Vision to Audio.}
Recent token pruning methods have emerged as a practical direction to reduce the sequence length~\cite{yang2025visionzip,liu2026globalcom2,liu2025vidcom2,chen2025v2drop,lin2026vcast,wang2025stc,han2026ficoco}. In the visual domain, FastV~\citep{chen2024image} ranks image tokens by their average attention scores received from text tokens, $\mathrm{importance}[k] = \frac{1}{H}\sum_h a_h[\text{last},k]$, and discards the lowest-scored ones. More recently, DART~\citep{wen-etal-2025-stop} argues that importance-based scoring has inherent positional bias and can sometimes underperform even random pruning. Instead, DART proposes a \emph{duplication-aware} metric that retains tokens with high cosine dissimilarity from a small set of pivot tokens, achieving strong results on image and vision understanding benchmarks with extreme reduction~\citep{wen-etal-2025-token}.

However, AudioMarathon~\citep{he2025audiomarathon} demonstrates that vision-centric pruning methods do not transfer well to audio. A key reason is that audio tasks are fundamentally divided into two distinct categories: \emph{semantic} tasks and \emph{acoustic} tasks. Each requiring attention to entirely different token-level features. Vision-domain methods, which are agnostic to this dichotomy, fail to preserve the task-critical tokens for either category, and both FastV and DART suffer significant performance degradation on long-audio tasks. In most cases, they perform worse than a simple Frame baseline that uniformly subsamples audio tokens, as shown in Table~\ref{tab:pruning_comparison_results_qwen}~\citep{he2025audiomarathon}.

%% file: section/4_method.tex
\section{Methodology}
\label{sec:method}

\subsection{Preliminaries}

\noindent \textbf{LALM Architecture.}
A typical Large Audio-Language Model (LALM) consists of three components: an audio encoder $g$, a projector $P$, and a pretrained large language model. Given an audio waveform $X_a$, the encoder produces a sequence of acoustic embeddings~\citep{qwen2.5-omni,ghosh2025audio,abouelenin2025phi}
\begin{equation}
Z_a = g(X_a) \in \mathbb{R}^{N_a \times d_e},
\end{equation}
which are mapped into the language model space by the projector,
\begin{equation}
H_a = P(Z_a).
\end{equation}
Given text tokens $H_q$, the model predicts the output sequence $Y$ autoregressively:
\begin{equation}
p(Y \mid H_a, H_q) = \prod_i p(y_i \mid H_a, H_q, y_{<i}).
\label{eq:lalm-decode}
\end{equation}

\noindent \textbf{Audio token pruning.}
Let the audio sequence contain $N_a$ tokens. Under pruning ratio $r$, the model retains
\begin{equation}
k = \lfloor N_a(1-r)\rfloor
\end{equation}
tokens and discards the rest. Given token importance scores:

$\mathrm{importance}[1], \ldots, \mathrm{importance}[N_a]$, the retained token set is
\begin{equation}
\mathcal{S} = \operatorname{TopK}\!\bigl(\mathrm{importance}[1], \ldots, \mathrm{importance}[N_a]; k\bigr),
\label{eq:token-select}
\end{equation}
where $\operatorname{TopK}(\cdot;k)$ returns the indices of the $k$ largest elements.

\subsection{Motivation}
\label{sec:motivation}

In LALMs, audio tokens dominate the input length and therefore dominate inference cost. A desirable pruning strategy should satisfy two properties. First, it should avoid positional bias, since informative audio content may appear anywhere in the sequence. Second, it should be content-adaptive, since speech, silence, background sound, and music do not contribute equally to downstream reasoning, while existing methods do not fully satisfy these requirements.

\noindent \textbf{Attention-based pruning suffers from positional bias.}
FastV estimates token importance from decoder attention:
\begin{equation}
\mathrm{importance}[k] = \frac{1}{N_h} \sum_{h=1}^{N_h} a_h[\mathrm{last}, k],
\label{eq:fastv}
\end{equation}
where $a_h[\mathrm{last}, k]$ is the attention from the last generated text token to audio token $k$ in head $h$. In audio settings, these scores are often biased toward the sequence tail and therefore do not reliably reflect acoustic salience.

\noindent \textbf{Uniform subsampling is content-blind.}
Frame-based pruning avoids positional bias by selecting tokens at regular temporal intervals. This preserves coarse coverage of the sequence, but it allocates the same budget to informative and uninformative regions. Under aggressive pruning, such content-blind allocation becomes especially harmful.

\begin{figure}[!b]
  \vspace{-4pt}
  \flushleft
  \includegraphics[width=0.9\linewidth]{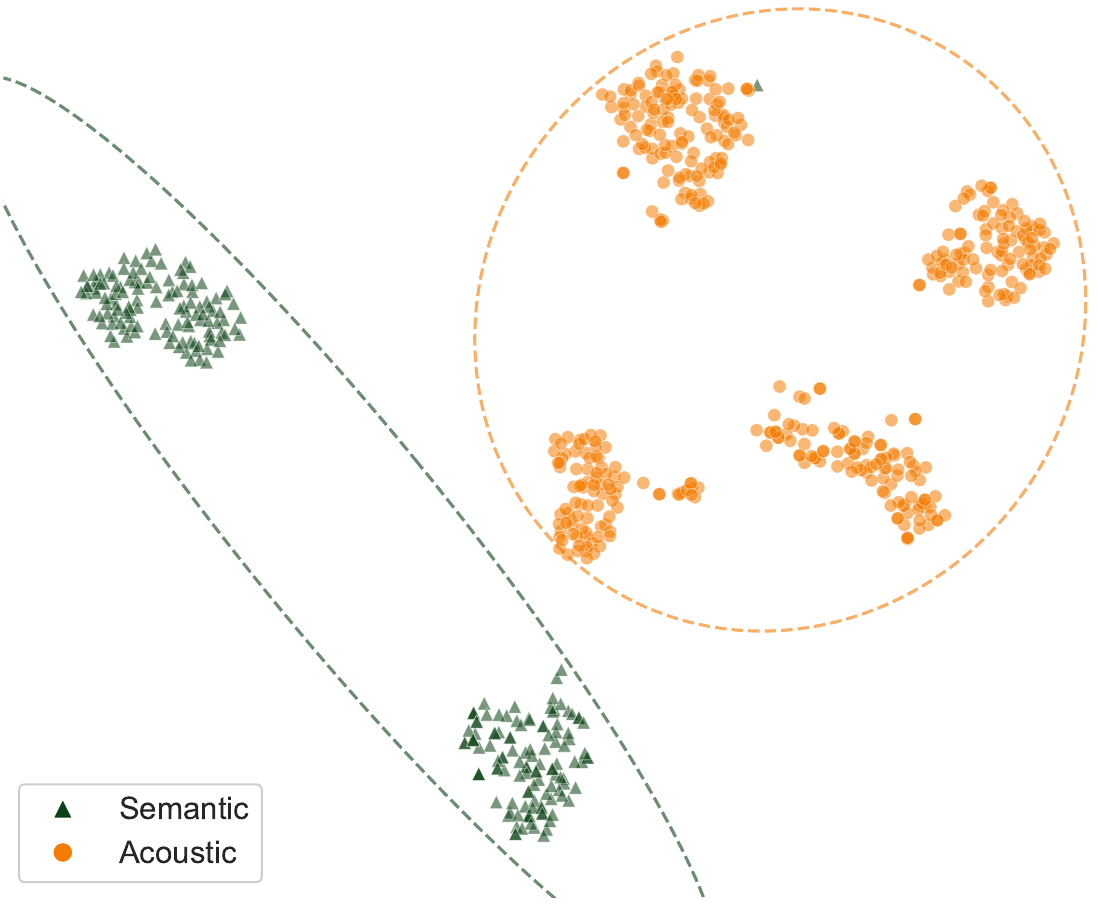}
  \caption{t-SNE visualization of per-sample head behavior. Each sample yields a 16-dimensional selectivity vector via Eq.~\eqref{eq:selectivity}). Semantic and acoustic tasks form clearly separated clusters, supporting the use of soft routing.}
  \label{fig:tsne}
  \Description{a figure showing different task types}
\end{figure}

Figure~\ref{fig:tsne} and Figure~\ref{fig:selectivity-heatmap} provide two complementary views of task-dependent head behavior. In the t-SNE visualization, semantic and acoustic tasks form clearly separated groups, while mixed tasks lie between them. The selectivity heatmap reveals the same trend at the head level: semantic tasks exhibit relatively diffuse and homogeneous selectivity across heads, whereas acoustic tasks rely more strongly on a smaller subset of highly selective heads, with mixed tasks again showing intermediate patterns. Together, these observations support our core motivation that heterogeneous audio tasks require different head-weight preferences, making a single fixed profile insufficient and motivating our routing-based formulation.

These observations suggest that effective audio token pruning should combine position-agnostic token scoring with input-adaptive head weighting.
\begin{figure}
    \centering
    \setlength{\abovecaptionskip}{4pt}
    \includegraphics[width=\linewidth]{figures/fig1c_selectivity_heatmap.png}
    \caption{\textbf{Selectivity heatmap across tasks and 16 heads of Qwen2.5-Omni-3B.} Semantic tasks show more diffuse and homogeneous selectivity, while acoustic tasks concentrate on a smaller subset of highly selective heads, illustrating \textit{head-behavior heterogeneity} on different tasks.}
    \vspace{-10pt}
    \label{fig:selectivity-heatmap}
    \Description{Heatmap showing head selectivity scores across AudioMarathon tasks, grouped by semantic, acoustic, and mixed categories.}
\end{figure}
\subsection{\ours{} Mechanism}
\label{sec:headrouter}

We propose \ours{}, a training-free pruning method that computes token importance from position-bias reduced text-to-audio probing and dynamically adjusts head weights according to the input. Let $N_h$ denote the total number of attention heads.

\subsubsection{Head-weighted token importance}
\label{sec:importance}

Given a per-head attention distribution $p_h[k]$ over audio tokens (defined in Section~\ref{sec:qk-probe}) and a head-weight vector $\mathbf{w} = (w_1, \ldots, w_{N_h})$, we normalize it as
\begin{equation}
\bar{w}_h = \frac{w_h}{\sum_{h'=1}^{N_h} w_{h'}}.
\label{eq:weight-norm}
\end{equation}
The importance of audio token $k$ is then computed by
\begin{equation}
\mathrm{importance}[k] = \sum_{h=1}^{N_h} \bar{w}_h \, p_h[k].
\label{eq:importance}
\end{equation}
Finally, the model retains the token subset defined by Eq.~\eqref{eq:token-select}.

\subsubsection{Task-specific head-weight profiles}
\label{sec:profiles}

Our method uses three fixed head-weight profiles. The semantic profile $\mathbf{w}^{\mathrm{sem}}$ is estimated offline from semantic tasks. The acoustic profile $\mathbf{w}^{\mathrm{aco}}$ is estimated offline from acoustic tasks. We also define a uniform weight profile
\begin{equation}
w_h^{\mathrm{uni}} = \frac{1}{N_h}, \qquad h = 1, \ldots, N_h.
\label{eq:uniform-weight}
\end{equation}
The semantic and acoustic profiles are obtained from offline head-ablation statistics. The uniform weight profile is defined analytically. No additional training is required.

\subsubsection{Per-head selectivity and routing signal}
\label{sec:selectivity}

To measure how concentrated each head is over audio tokens, \textit{selectivity} is defined:
\begin{equation}
\mathrm{sel}_h
= 1 - \frac{\mathcal{H}(p_h)}{\log N_a}
= 1 + \frac{1}{\log N_a}\sum_{k=1}^{N_a} p_h[k]\log p_h[k].
\label{eq:selectivity}
\end{equation}
Here $\mathcal{H}(p_h) = -\sum_k p_h[k]\log p_h[k]$ is the entropy of the head-wise audio attention distribution. By construction, $\mathrm{sel}_h \in [0,1]$. A larger value indicates that head $h$ is more selective, while a smaller value indicates a more diffuse distribution.

We then summarize the diversity of head behaviors by the standard deviation of selectivity across heads:
\begin{equation}
\mathrm{spr} = \operatorname{std}\!\left([\mathrm{sel}_1, \mathrm{sel}_2, \ldots, \mathrm{sel}_{N_h}]\right).
\label{eq:spread}
\end{equation}
A small $\mathrm{spr}$ indicates that the heads behave similarly, which is more typical of semantic inputs. A large $\mathrm{spr}$ indicates that only a subset of heads is highly selective, which is more typical of acoustic inputs.

\subsubsection{Gaussian soft routing}
\label{sec:gaussian-routing}

We use the routing signal in Eq.~\eqref{eq:spread} to softly combine the three head-weight profiles. Let the profile be
\begin{equation}
\mathcal{W} = \left\{
\mathbf{w}^{\mathrm{sem}},
\mathbf{w}^{\mathrm{uni}},
\mathbf{w}^{\mathrm{aco}}
\right\},
\end{equation}
with corresponding centers
$\mu_{\mathrm{sem}}$, $\mu_{\mathrm{uni}}$,  $\mu_{\mathrm{aco}}$.

For an input with routing signal $\mathrm{spr}$, we compute the mixing coefficient of profile $c \in \{\mathrm{sem}, \mathrm{uni}, \mathrm{aco}\}$ by
\begin{equation}
\alpha_c
=
\frac{
\exp\!\left(-\frac{(\mathrm{spr}-\mu_c)^2}{2\sigma_G^2}\right)
}{
\sum_{c'} \exp\!\left(-\frac{(\mathrm{spr}-\mu_{c'})^2}{2\sigma_G^2}\right)
},
\label{eq:alpha}
\end{equation}
where $\sigma_G$ is the bandwidth of the Gaussian kernel. The final head-weight vector is
\begin{equation}
\mathbf{w}^{*}
=
\alpha_{\mathrm{sem}} \mathbf{w}^{\mathrm{sem}}
+
\alpha_{\mathrm{uni}} \mathbf{w}^{\mathrm{uni}}
+
\alpha_{\mathrm{aco}} \mathbf{w}^{\mathrm{aco}}.
\label{eq:wstar}
\end{equation}
After normalization, $\mathbf{w}^{*}$ is used in Eq.~\eqref{eq:importance}.

Compared with hard threshold routing, this formulation has three advantages. It preserves continuity near profile boundaries, reduces sensitivity to threshold selection, and naturally interpolates mixed inputs that do not belong to a single category.

\subsubsection{Position-agnostic QK probing}
\label{sec:qk-probe}

The per-head attention distributions $p_h[k]$ used in Eq.~\eqref{eq:importance} and Eq.~\eqref{eq:selectivity} are obtained by probing text-to-audio interactions without positional encoding. Let $M$ denote the pruning layer. We use the query and key projections from layer $M-1$. For each attention head $h$, we compute
\begin{equation}
Q_h = H_{\mathrm{text}} W_h^Q, \qquad
K_h = H_{\mathrm{audio}} W_h^K,
\label{eq:qk-proj}
\end{equation}
where no RoPE is applied during this probing step. The head-wise text-to-audio affinity is
\begin{equation}
A_h[j,k] = \frac{Q_h^{(j)} K_h^{(k)\top}}{\sqrt{d_k}},
\label{eq:affinity}
\end{equation}
where $j$ indexes text tokens and $k$ indexes audio tokens. We then average the normalized attention over all text tokens to obtain a marginal attention distribution over audio tokens:
\begin{equation}
p_h[k] = \frac{1}{N_t} \sum_{j=1}^{N_t} \operatorname{softmax}_k\!\bigl(A_h[j,:]\bigr)[k].
\label{eq:marginal-attn}
\end{equation}
By removing positional encoding from this probing step, the resulting score depends purely on semantic alignment between text and audio content rather than on absolute token position, which avoids the positional bias observed in attention-based methods like FastV.

\subsubsection{Complexity}
\label{sec:complexity}

The additional cost of routing is small. Beyond the QK probe already used for token scoring, the routing module only requires one entropy computation per head, one standard deviation over $N_h$ values, and a three-way weighted combination of fixed profiles. Its computational complexity is therefore negligible compared with the overall prefilling pass. Empirically, our profiling results further show that the routing step accounts for less than 1\% of the total prefilling time, confirming that the proposed mechanism introduces only minimal runtime overhead.

{
\setlength{\textfloatsep}{6pt}
\begin{algorithm}[t]
\caption{\ours{} inference pipeline}
\label{alg:headrouter}
\begin{algorithmic}[1]
\REQUIRE Audio input, text input, pruning ratio $r$, three head-weight profiles, profile centers, bandwidth $\sigma_G$
\STATE Encode the audio and obtain audio tokens
\STATE Embed the text tokens
\STATE Forward the full sequence through layers $0$ to $M-1$
\STATE Compute position-agnostic text-to-audio probe scores using Eq.~\eqref{eq:qk-proj} to Eq.~\eqref{eq:marginal-attn}
\STATE Compute per-head selectivity by Eq.~\eqref{eq:selectivity}
\STATE Compute the routing signal by Eq.~\eqref{eq:spread}
\STATE Compute the soft routing coefficients by Eq.~\eqref{eq:alpha}
\STATE Obtain the final head-weight vector by Eq.~\eqref{eq:wstar}
\STATE Compute token importance by Eq.~\eqref{eq:importance}
\STATE Retain the top $\lfloor N_a(1-r)\rfloor$ audio tokens
\STATE Continue the forward pass on the pruned sequence
\end{algorithmic}
\end{algorithm}
}

%% file: section/5_experiments.tex
\input{tables/tab_main_results}
\input{tables/tab_mmaupro_results}
\section{Experiments}
\label{sec:experiments}

\subsection{Experimental Setup}
\textbf{Evaluation benchmarks.} We evaluate on two complementary benchmarks. (1)~AudioMarathon~\citep{he2025audiomarathon} covers a diverse set of audio understanding tasks. Following our analysis setup, we group the tasks discussed in this work into three categories: semantic tasks, including Automatic Speech Recognition (ASR) and Speech Entity Recognition (SER); acoustic tasks, including Music Classifier (MC), Audio Scene Classifier (ASC), Sound Event Detection (SED), Emotion Recognition (ER), Speaker Age Recognition (SAR), and Speaker Gender Recognition (SGR). (2)~MMAU-Pro~\citep{kumar2025mmau} provides a challenging multiple-choice benchmark for broad audio understanding. It spans five audio domains, namely Speech, Voice Chat, Sound, Music, and Spatial Audio, and contains a total of 2{,}567 questions. For the purpose of analysis, we further group these domains into two broader categories: a semantic group, consisting of Speech and Voice Chat (VC), and an acoustic group, consisting of Sound, Music, and Spatial Audio (SA). Under this partition, MMAU-Pro contains 751 semantic questions and 1{,}816 acoustic questions.

\noindent \textbf{Comparison methods and models.} We evaluate at pruning ratios r$ \in \{0.3, 0.6, 0.9\}$ and compare \ours{} against four competitive baselines: the unpruned full model, Random token pruning, FastV~\citep{chen2024image}, DART~\citep{wen-etal-2025-stop}, and Frame~\citep{he2025audiomarathon}. Our main experiments are conducted on Qwen2.5-Omni-3B~\citep{qwen2.5-omni}. We further report results on Qwen2.5-Omni-7B~\citep{qwen2.5-omni} and Phi-4-Multimodal~\citep{abouelenin2025phi} to evaluate cross-model transfer. For analysis, we also consider fixed-profile variants based on $\mathbf{w}^{\mathrm{sem}}$ and $\mathbf{w}^{\mathrm{aco}}$, whose results are shown in~\ref{fig:ablation_bar}.

\noindent \textbf{Implementation setup.} Following common practice in audio token-pruning pipelines, we apply pruning at the second audio layer ($M{=}2$). In preliminary experiments, pruning at the first layer caused substantially larger performance drops, suggesting that the earliest audio representations are still too fragile for aggressive compression. \ours{} uses the no-RoPE QK probe from the preceding layer $M{-}1$ for both token importance scoring and routing. The Gaussian kernel centers $\mu_k$ are estimated from calibration statistics collected on a small held-out set, using only 10 samples per category from the AudioMarathon development split. The bandwidth $\sigma_G$ is selected from the same calibration procedure. Notably, our experiments show that this lightweight calibration is sufficient: increasing the calibration set to 50\% of the development split changes the final routing parameters by less than 5\%, confirming that the selectivity statistics converge quickly with very few samples. For evaluation metrics, we report \textbf{F1-score} on all multiple-choice and classification tasks to account for potential class imbalance, and \textbf{Word Accuracy Rate} (WAR $= 1 - \text{Word Error Rate (WER)}$) for ASR, so that higher values uniformly indicate better performance across all tasks. Due to the category imbalance in MMAU-Pro, the overall score is computed as a \textbf{sample-weighted average}.

\subsection{Main Results}
\label{sec:main-results}
\input{tables/tab_cross_model_mmaupro}
Tables~\ref{tab:pruning_comparison_results_qwen} and~\ref{tab:pruning_comparison_results_mmaupro} highlight three consistent trends. (i) On AudioMarathon, \textbf{\ours{}} is the only method that remains the strongest on average across all three pruning ratios on Qwen2.5-Omni-3B, with the clearest margin appearing at medium pruning, where it outperforms the strongest baseline by 2.5 points. (ii) This advantage does not come from a single task: even when DART remains stronger on ASR at extreme pruning, \ours{} preserves a better balance across semantic and acoustic tasks, which leads to the best overall average. (iii) On MMAU-Pro, the same pattern holds: the drop from the unpruned model stays very small at light and medium pruning, and \ours{} still remains ahead of the best baseline by more than 1 point under the most aggressive pruning setting.
Taken together, these results support our main claim that adaptive head weighting is especially beneficial when token budgets become tight and a single fixed profile is no longer reliable.

Figure~\ref{fig:oracle-r06} further compares \ours{} with an oracle reference at pruning ratio r$=0.6$. Here, the oracle reflects token selection that is better aligned with naturally salient high-energy audio segments. \ours{} remains consistently closer to this reference than Frame and FastV across the reported tasks. This result suggests that the proposed method captures informative audio regions more effectively than purely time-uniform selection in Frame and is substantially less affected by the positional bias that limits FastV.

\noindent \textbf{Cross-benchmark generalization on MMAU-Pro.}
The results on MMAU-Pro further show that our method does not only overfit to AudioMarathon. In particular, \ours{} stays nearly lossless up when pruning 60\% of tokens on Qwen2.5-Omni-3B, while preserving a clear advantage over Frame and FastV. Even when pruning 90\% of tokens, where all methods degrade substantially, \ours{} still keeps the best overall balance. This suggests that the routing signal captures a fairly general distinction between semantic and acoustic audio inputs across different benchmarks, demonstrating the consistent robustness of \ours{}.

\noindent \textbf{Cross-model generalization.}
Notably, \ours{} is especially strong on ASR, a task that is highly sensitive to token selection errors: on Qwen2.5-Omni-3B at r${=}0.6$, it improves ASR performance by 6.6 points over Frame. More broadly, this advantage transfers across model families. On AudioMarathon, \ours{} remains the best overall method on both Qwen2.5-Omni-7B and Phi-4-Multimodal, outperforming the strongest baseline by 1.2--2.4 points and 0.4--2.0 points across pruning ratios, respectively. On MMAU-Pro, the trend is similarly stable for Phi-4-Multimodal, where \ours{} leads by 0.5--0.9 points, while the ranking on Qwen2.5-Omni-7B is tighter. Overall, these results indicate that the benefit of routing is not tied to a single backbone.
\begin{figure}[t]

  \centering
  \includegraphics[width=\linewidth]{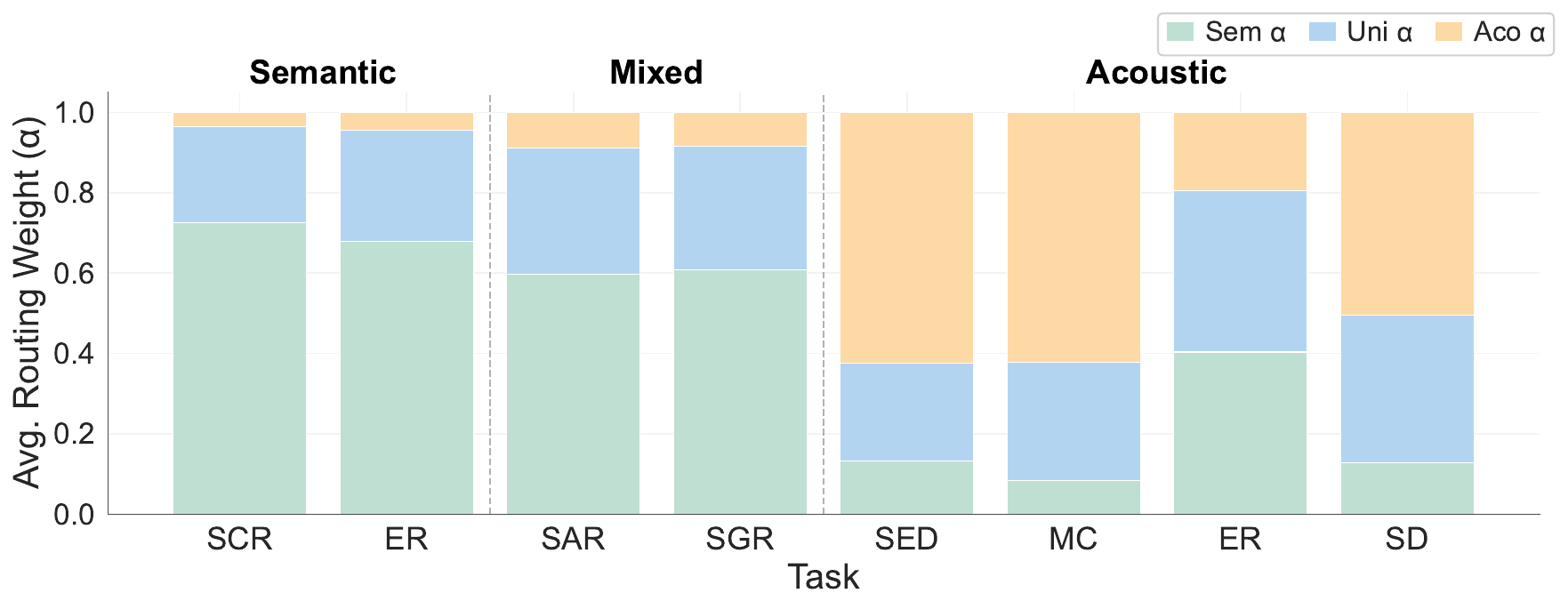}
  \vspace{-9mm}
  \caption{Dynamic routing distribution. Average mixing coefficients ($\bar{\alpha}_k$) per task. Stacked components represent Semantic, Uniform, and Acoustic profiles, adaptively assigning target profiles.}
  \vspace{-3mm}
  \label{fig:routing-dist}
  \Description{Stacked bar chart showing routing coefficient distribution across AudioMarathon tasks.}
\end{figure}
\input{figures/fig_efficiency}
\input{figures/fig_ablation_bar}
\subsection{Routing Distribution}
\label{sec:routing-accuracy}

Figure~\ref{fig:routing-dist} visualizes the average Gaussian routing coefficients $\alpha_k$ across all AudioMarathon tasks, where each stacked bar shows how the router distributes mass among the semantic, uniform, and acoustic profiles for a given task. A higher $\alpha_{\text{sem}}$ component indicates the router judges the input as semantically dominated; a higher $\alpha_{\text{aco}}$ component indicates acoustically dominated behavior.

Two patterns emerge clearly. First, the routing is well-aligned with task semantics: speech-related tasks such as ASR, SER, and SCR are predominantly routed toward $\mathbf{w}^{\mathrm{sem}}$, while acoustic tasks such as SGR, SED, MC, and SAR are routed toward $\mathbf{w}^{\mathrm{aco}}$. This confirms that the selectivity spread $\mathrm{spr}$ introduced in Section~\ref{sec:selectivity} captures the semantic-vs-acoustic distinction effectively from the head-behavior statistics alone, without task labels at inference time.

Second, mixed-nature tasks like Emotion Recognition (ER) receive non-trivial mass on all three profiles simultaneously, producing blended head weights that a hard-threshold assignment cannot express. This continuous interpolation is precisely the advantage of the Gaussian soft routing formulation: instead of committing to a single profile, the router smoothly adjusts the head-weight vector to match the actual head-behavior pattern of each input.

\subsection{Ablation Study}
\label{sec:ablation}

We ablate three key components of \ours{} under different token pruning ratios: the routing strategy, the routing module itself, and the frame-based downsampling stage. The full \ours{} combines Gaussian soft routing, the Frame pre-filter, and the no-RoPE probe. \textit{Hard Threshold} replaces Gaussian soft routing with hard threshold-based discrete profile assignment while keeping the rest of the pipeline unchanged. \textit{w/o Router} removes the routing module entirely, applying uniform head weights with the Frame pre-filter and no-RoPE probe. \textit{w/o Downsampling} removes the Frame pre-filter and relies on routing alone for token selection. This design allows us to isolate the contribution of each component in Figure~\ref{fig:ablation_bar}.

Figure~\ref{fig:ablation_bar} shows three clear trends on both AudioMarathon and MMAU-Pro. First, the full model consistently performs best or remains the most stable across pruning ratios, indicating that the three components are complementary rather than redundant. Second, comparing \ours{} with \textit{Hard Threshold} shows that Gaussian soft routing is more reliable than discrete assignment, especially at higher pruning ratios where hard thresholding becomes more brittle. Third, comparing \ours{} with \textit{w/o Router} confirms that the routing module provides a meaningful gain beyond the no-RoPE probe alone, particularly under aggressive compression. Finally, \textit{w/o Downsampling}, which bypasses the frame-based pre-filter, falls clearly behind \ours{}, demonstrating that the two-stage pipeline—first reducing via temporal downsampling, then refining via adaptive scoring—yields better robustness than routing alone.

\subsection{Memory and Performance Trade-off}
\label{sec:efficiency}

Figure~\ref{fig:efficiency} plots F1-score against peak GPU memory across five AudioMarathon tasks at pruning ratios r$ \in \{0.3, 0.4, \ldots, 0.9\}$. \ours{} consistently occupies the Pareto-optimal region, achieving higher task performance at comparable memory cost ($\sim$10\,GB, same regime as Frame and DART). FastV is omitted due to its much higher memory ceiling (${\sim}15$--$20$\,GB). As the pruning ratio increases, the advantage of \ours{} becomes more pronounced, since adaptive head weighting retains more task-relevant information within the same memory envelope. A similar trend is observed in inference latency: at r${=}0.6$, \ours{} reduces LibriSpeech latency from 102.3\,s to 41.9\,s while improving the average score by 2.5 points, confirming a favorable performance--efficiency trade-off under aggressive compression.

%% file: tables/tab_main_results.tex
\begin{table*}[!t]

\caption{Cross-model performance on AudioMarathon with Qwen2.5-Omni-3B, Qwen2.5-Omni-7B, and Phi-4-Multimodal across token pruning methods at three retention ratios. F1-score (0--100) is reported for all tasks except ASR, which is Word Accuracy Rate (WAR). Vanilla Avg.\ set to 100\%; other Avg.\ shown as \% of Vanilla. Best Avg.\ per ratio in \textbf{bold}.}

\centering
\setlength{\tabcolsep}{3.5pt}
\renewcommand{\arraystretch}{0.6}
\resizebox{\textwidth}{!}{%
\begin{tabular}{w{l}{3cm}|w{c}{1.0cm}w{c}{1.0cm}w{c}{1.0cm}w{c}{1.0cm}w{c}{1.0cm}w{c}{1.0cm}w{c}{1.0cm}w{c}{1.0cm}w{c}{1.0cm}|w{c}{1.5cm}|w{c}{1.5cm}}
\toprule[1.5pt]
\multirow{2}{*}{\textbf{Method}}
& \multicolumn{3}{c}{\textbf{Semantic}}
& \multicolumn{6}{c|}{\textbf{Acoustic}}
& \multirow{2}{*}{\textbf{Avg.}} & \multirow{2}{*}{\textbf{Diff.}} \\
\cmidrule(lr){2-4} \cmidrule(lr){5-10}
& $SER$ & $SCR$ & $ASR$ & $SED$ & $MC$ & $ASC$ & $ER$ & $SAR$ & $SGR$ & & \\ \midrule

\rowcolor{gray!12}
Qwen2.5-Omni-3B & \multicolumn{11}{c}{\textit{Upper Bound, All Tokens (100\%)}} \\
Vanilla & 25.2 & 82.3 & {94.7} & {70.2} & {97.4} & {69.3} & 39.6 & 29.1 & 97.2 & 100.0\% & -- \\
\rowcolor{gray!12}
Qwen2.5-Omni-3B & \multicolumn{11}{c}{\textit{Retain 70\% Tokens ($\downarrow$ 30\%)}} \\
Random & 26.5 & 80.3 & 88.4 & 71.1 & 97.5 & 69.7 & \textbf{38.4} & 28.6 & 95.7 & 98.5\% & $-1.5\%$ \\
FastV~\scriptsize{(ECCV24)} & 18.7 & 68.2 & 76.3 & 61.3 & 98.4 & 57.2 & 31.1 & 17.3 & 97.5 & 86.9\% & $-13.1\%$ \\
DART~\scriptsize{(EMNLP25)} & 23.2 & 74.2 & 81.4 & 73.1 & 97.6 & \textbf{72.5} & 37.1 & 23.0 & 48.7 & 87.8\% & $-12.2\%$ \\
Frame~\scriptsize{(2026.1)} & 26.8 & 80.9 & 92.2 & 70.5 & 98.5 & 70.2 & 36.4 & \textbf{31.4} & 96.7 & 99.9\% & $-0.1\%$ \\
\textbf{\ours{}} & \textbf{28.8} & \textbf{82.9} & \textbf{96.8} & \textbf{73.6} & \textbf{100.0} & 69.4 & 34.3 & 30.6 & \textbf{99.4} & \textbf{101.8\%} & $\textbf{+1.8\%}$ \\
\rowcolor{gray!12}
Qwen2.5-Omni-3B & \multicolumn{11}{c}{\textit{Retain 40\% Tokens ($\downarrow$ 60\%)}} \\
Random & 24.2 & 75.3 & 59.7 & 68.7 & 95.8 & 68.3 & 37.9 & 27.2 & 93.5 & 91.1\% & $-8.9\%$ \\
FastV~\scriptsize{(ECCV24)} & 18.0 & 63.8 & 39.2 & 60.5 & 97.5 & 57.8 & 28.6 & 17.1 & 95.3 & 79.0\% & $-21.0\%$ \\
DART~\scriptsize{(EMNLP25)} & 23.1 & 64.6 & 62.8 & 71.9 & 99.1 & \textbf{73.4} & 37.6 & 28.1 & 46.0 & 83.8\% & $-16.2\%$ \\
Frame~\scriptsize{(2026.1)} & 25.6 & 75.3 & 82.2 & 69.0 & \textbf{100.0} & 68.3 & \textbf{38.6} & 28.3 & 91.0 & 95.7\% & $-4.3\%$ \\
\textbf{\ours{}} & \textbf{29.6} & \textbf{81.8} & \textbf{88.8} & \textbf{72.8} & \textbf{100.0} & 67.8 & 30.9 & \textbf{30.2} & \textbf{99.6} & \textbf{99.4\%} & $\textbf{-0.6\%}$ \\
\rowcolor{gray!12}
Qwen2.5-Omni-3B & \multicolumn{11}{c}{\textit{Retain 10\% Tokens ($\downarrow$ 90\%)}} \\
Random & 24.0 & 58.1 & 0.0 & 65.9 & 97.6 & 60.0 & 41.8 & 17.1 & 84.3 & 74.3\% & $-25.7\%$ \\
FastV~\scriptsize{(ECCV24)} & 16.8 & 54.9 & 3.5 & 65.2 & 95.9 & 55.9 & 32.7 & 14.8 & 86.5 & 70.5\% & $-29.5\%$ \\
DART~\scriptsize{(EMNLP25)} & 17.3 & 54.1 & \textbf{62.9} & 66.8 & 99.1 & \textbf{69.3} & \textbf{42.6} & 22.1 & 52.8 & 80.5\% & $-19.5\%$ \\
Frame~\scriptsize{(2026.1)} & 22.8 & 58.2 & 0.0 & 64.5 & 95.0 & 60.9 & 41.1 & 18.1 & 87.0 & 74.0\% & $-26.0\%$ \\
\textbf{\ours{}} & \textbf{29.2} & \textbf{63.0} & 5.4 & \textbf{67.7} & \textbf{99.2} & 63.2 & 34.3 & \textbf{29.3} & \textbf{96.5} & \textbf{80.7\%} & $\textbf{-19.3\%}$ \\
\midrule

\rowcolor{gray!12}
Qwen2.5-Omni-7B & \multicolumn{11}{c}{\textit{Upper Bound, All Tokens (100\%)}} \\
Vanilla & 26.3 & 85.1 & {98.1} & {78.4} & {100.0} & {72.2} & 53.4 & 21.4 & 98.0 & 100.0\% & -- \\
\rowcolor{gray!12}
Qwen2.5-Omni-7B & \multicolumn{11}{c}{\textit{Retain 70\% Tokens ($\downarrow$ 30\%)}} \\
Random & 27.4 & 87.0 & 91.5 & 78.7 & 98.3 & 72.5 & 44.4 & 30.3 & \textbf{99.5} & 99.4\% & $-0.6\%$ \\
FastV~\scriptsize{(ECCV24)} & \textbf{27.6} & 83.4 & 55.3 & 77.6 & 99.2 & 73.0 & \textbf{47.2} & 27.4 & 99.4 & 93.3\% & $-6.7\%$ \\
DART~\scriptsize{(EMNLP25)} & 26.5 & 84.0 & 86.2 & 79.1 & 99.2 & \textbf{73.6} & 44.9 & 29.0 & 99.3 & 98.3\% & $-1.7\%$ \\
Frame~\scriptsize{(2026.1)} & \textbf{27.6} & \textbf{87.1} & 94.9 & \textbf{79.5} & 99.2 & 71.7 & 41.6 & 29.2 & \textbf{99.5} & 99.6\% & $-0.4\%$ \\
\textbf{\ours{}} & 25.9 & 86.8 & \textbf{95.2} & 78.0 & \textbf{99.2} & 73.2 & 43.3 & \textbf{50.9} & \textbf{99.5} & \textbf{103.0\%} & $\textbf{+3.0\%}$ \\
\rowcolor{gray!12}
Qwen2.5-Omni-7B & \multicolumn{11}{c}{\textit{Retain 40\% Tokens ($\downarrow$ 60\%)}} \\
Random & 26.5 & 81.8 & 69.3 & 77.2 & 99.2 & 72.5 & 44.9 & 28.3 & 99.3 & 94.7\% & $-5.3\%$ \\
FastV~\scriptsize{(ECCV24)} & 25.7 & 71.8 & 0.0 & 78.7 & 98.3 & 71.0 & 51.1 & 25.4 & 99.3 & 82.4\% & $-17.6\%$ \\
DART~\scriptsize{(EMNLP25)} & \textbf{29.0} & 76.2 & 46.6 & 76.4 & 99.2 & \textbf{73.5} & \textbf{47.2} & 24.3 & 97.9 & 90.2\% & $-9.8\%$ \\
Frame~\scriptsize{(2026.1)} & 27.6 & \textbf{83.8} & 82.2 & \textbf{78.4} & \textbf{99.2} & 71.3 & 39.3 & 25.9 & \textbf{99.3} & 95.9\% & $-4.1\%$ \\
\textbf{\ours{}} & 28.0 & 80.4 & \textbf{64.3} & \textbf{78.7} & \textbf{99.2} & 72.2 & 43.8 & \textbf{51.0} & \textbf{99.7} & \textbf{97.6\%} & $\textbf{-2.4\%}$ \\
\rowcolor{gray!12}
Qwen2.5-Omni-7B & \multicolumn{11}{c}{\textit{Retain 10\% Tokens ($\downarrow$ 90\%)}} \\
Random & 28.0 & \textbf{65.5} & 0.0 & 76.4 & \textbf{99.2} & 68.9 & 44.4 & 27.1 & \textbf{99.4} & 80.4\% & $-19.6\%$ \\
FastV~\scriptsize{(ECCV24)} & 25.3 & 64.3 & 1.8 & 75.2 & 96.7 & 64.5 & 46.1 & 30.6 & 99.3 & 79.7\% & $-20.3\%$ \\
DART~\scriptsize{(EMNLP25)} & \textbf{31.6} & \textbf{65.5} & 0.0 & 71.3 & 98.3 & 65.3 & 41.6 & 20.5 & 73.4 & 73.8\% & $-26.2\%$ \\
Frame~\scriptsize{(2026.1)} & 27.4 & 66.1 & 0.0 & 73.2 & \textbf{99.2} & 66.2 & 43.3 & 24.8 & 99.3 & 78.9\% & $-21.1\%$ \\
\textbf{\ours{}} & 26.9 & 64.2 & \textbf{0.0} & \textbf{72.8} & \textbf{99.2} & \textbf{68.1} & \textbf{46.1} & \textbf{49.0} & 99.1 & \textbf{83.1\%} & $\textbf{-16.9\%}$ \\
\midrule

\rowcolor{gray!12}
Phi-4-Multimodal & \multicolumn{11}{c}{\textit{Upper Bound, All Tokens (100\%)}} \\
Vanilla & 18.4 & 69.3 & {92.7} & 55.1 & 46.7 & 23.4 & 27.3 & 26.6 & 91.1 & 100.0\% & -- \\
\rowcolor{gray!12}
Phi-4-Multimodal & \multicolumn{11}{c}{\textit{Retain 70\% Tokens ($\downarrow$ 30\%)}} \\
Random & 18.4 & 67.5 & 49.1 & 31.4 & 39.8 & 30.2 & \textbf{31.0} & 24.5 & 93.6 & 85.4\% & $-14.6\%$ \\
FastV~\scriptsize{(ECCV24)} & 18.3 & 64.0 & 43.9 & 33.2 & 40.6 & 29.6 & 29.1 & 25.7 & 92.9 & 83.6\% & $-16.4\%$ \\
DART~\scriptsize{(EMNLP25)} & 16.8 & \textbf{67.6} & 57.2 & \textbf{54.5} & \textbf{46.1} & \textbf{31.8} & 28.6 & 27.1 & 91.6 & 93.4\% & $-6.6\%$ \\
Frame~\scriptsize{(2026.1)} & 17.7 & 64.4 & \textbf{63.4} & 31.4 & 32.6 & 29.0 & \textbf{31.0} & 27.4 & 92.4 & 86.4\% & $-13.6\%$ \\
\textbf{\ours{}} & \textbf{24.7} & 67.0 & 51.0 & 53.5 & 42.5 & 29.4 & 29.2 & \textbf{32.8} & \textbf{94.5} & \textbf{94.2\%} & $\textbf{-5.8\%}$ \\
\rowcolor{gray!12}
Phi-4-Multimodal & \multicolumn{11}{c}{\textit{Retain 40\% Tokens ($\downarrow$ 60\%)}} \\
Random & 18.7 & 61.6 & 7.9 & 30.3 & 27.4 & \textbf{30.6} & 29.4 & 20.5 & 91.2 & 70.5\% & $-29.5\%$ \\
FastV~\scriptsize{(ECCV24)} & 26.1 & 52.8 & 0.0 & 32.5 & 28.2 & 30.3 & 28.0 & 22.2 & 89.4 & 68.7\% & $-31.3\%$ \\
DART~\scriptsize{(EMNLP25)} & 18.0 & 61.1 & \textbf{23.7} & \textbf{53.9} & \textbf{44.8} & 25.4 & 29.4 & 24.4 & 88.0 & 81.8\% & $-18.2\%$ \\
Frame~\scriptsize{(2026.1)} & 23.8 & 59.0 & 23.3 & 31.1 & 28.5 & 30.0 & \textbf{30.1} & 22.3 & 87.8 & 74.5\% & $-25.5\%$ \\
\textbf{\ours{}} & \textbf{26.9} & \textbf{61.8} & 14.6 & 52.0 & 44.2 & 29.6 & 29.8 & \textbf{29.1} & \textbf{91.3} & \textbf{84.0\%} & $\textbf{-16.0\%}$ \\
\rowcolor{gray!12}
Phi-4-Multimodal & \multicolumn{11}{c}{\textit{Retain 10\% Tokens ($\downarrow$ 90\%)}} \\
Random & 18.7 & 35.3 & 0.0 & 29.3 & 20.6 & 29.6 & \textbf{33.4} & 11.1 & 67.5 & 54.5\% & $-45.5\%$ \\
FastV~\scriptsize{(ECCV24)} & 23.4 & 43.0 & 0.0 & 27.6 & 30.1 & 29.3 & 24.9 & 17.3 & \textbf{82.9} & 61.7\% & $-38.3\%$ \\
DART~\scriptsize{(EMNLP25)} & 16.8 & 49.3 & 0.0 & \textbf{52.0} & 40.2 & 24.4 & 27.4 & 18.6 & 77.2 & 67.9\% & $-32.1\%$ \\
Frame~\scriptsize{(2026.1)} & 24.6 & 36.8 & 0.0 & 28.4 & 25.0 & 28.1 & 30.1 & 12.1 & 66.6 & 55.9\% & $-44.1\%$ \\
\textbf{\ours{}} & \textbf{25.7} & \textbf{51.5} & \textbf{0.2} & 44.5 & \textbf{40.8} & \textbf{29.8} & 28.7 & \textbf{29.4} & 73.8 & \textbf{71.9\%} & $\textbf{-28.1\%}$ \\
\bottomrule[1.5pt]
\end{tabular}%
}
\label{tab:pruning_comparison_results_qwen}
\end{table*}

%% file: tables/tab_mmaupro_results.tex
\begin{table*}[!t]

\caption{Comparative experiments on MMAU-Pro with Qwen2.5-Omni-3B and Qwen2.5-Omni-7B models. Performance of vanilla Avg.\ set to 100\%; other Avg.\ shown as \% of Vanilla. Best Avg.\ per ratio in \textbf{bold}.}

\centering
\setlength{\tabcolsep}{4pt}
\renewcommand{\arraystretch}{0.85}
\resizebox{0.95\textwidth}{!}{%
\begin{tabular}{lccccc|cc|lccccc|cc}
\toprule[1.5pt]
\multicolumn{8}{c|}{\textbf{Qwen2.5-Omni-3B}} & \multicolumn{8}{c}{\textbf{Qwen2.5-Omni-7B}} \\
\midrule
\multirow{2}{*}{\textbf{Method}}
& \multicolumn{2}{c}{\textbf{Speech}}
& \multicolumn{3}{c|}{\textbf{Acoustic}}
& \multirow{2}{*}{\textbf{Avg.}} & \multirow{2}{*}{\textbf{Diff.}}
& \multirow{2}{*}{\textbf{Method}}
& \multicolumn{2}{c}{\textbf{Speech}}
& \multicolumn{3}{c|}{\textbf{Acoustic}}
& \multirow{2}{*}{\textbf{Avg.}} & \multirow{2}{*}{\textbf{Diff.}} \\
\cmidrule(lr){2-3} \cmidrule(lr){4-6} \cmidrule(lr){10-11} \cmidrule(lr){12-14}
    & $Speech$ & $Voice Chat$ & $Sound$ & $Music$ & $Spatial$ & &
    & & $Speech$ & $Vocie Chat$ & $Sound$ & $Music$ & $Spatial$ & & \\ \midrule

\rowcolor{gray!20}
\multicolumn{8}{c|}{\textit{Vanilla}} & \multicolumn{8}{c}{\textit{Vanilla}} \\
& 59.0 & 54.1 & 50.6 & 65.5 & 46.0 & 100.0\% & --
& & 60.6 & 63.6 & 57.5 & 64.8 & 29.7 & 100.0\% & -- \\
\rowcolor{mygray}
\multicolumn{8}{c|}{{\textit{Light Token Pruning} \ $\fg{(\downarrow 30\%)}$}}
& \multicolumn{8}{c}{{\textit{Light Token Pruning} \ $\fg{(\downarrow 30\%)}$}} \\
FastV~\scriptsize{(ECCV24)}
& 52.5 & 47.0 & 42.4 & 58.6 & \textbf{51.0} & 91.5\% & $-8.5\%$
& FastV~\scriptsize{(ECCV24)} & 58.0 & \textbf{64.1} & 56.5 & 64.3 & 27.5 & 97.2\% & $-2.8\%$ \\
DART~\scriptsize{(EMNLP25)}
& 55.6 & 50.3 & 49.5 & 64.4 & 44.0 & 96.0\% & $-4.0\%$
& DART~\scriptsize{(EMNLP25)} & 58.6 & 63.9 & 58.1 & \textbf{65.4} & 28.0 & \textbf{98.7\%} & $\textbf{-1.3\%}$ \\
Frame~\scriptsize{(2026.1)}
& 57.7 & \textbf{54.6} & 50.6 & 64.9 & 44.5 & 99.1\% & $-0.9\%$
& Frame~\scriptsize{(2026.1)} & 58.1 & 61.8 & \textbf{59.2} & 65.2 & 26.0 & 98.4\% & $-1.6\%$ \\
\textbf{\ours{}}
& \textbf{57.9} & \textbf{54.6} & \textbf{51.2} & \textbf{65.2} & 45.0 & \textbf{99.6\%} & $\textbf{-0.4\%}$
& \textbf{\ours{}} & \textbf{59.0} & 61.8 & 58.1 & 65.2 & \textbf{29.0} & \textbf{98.7\%} & $\textbf{-1.3\%}$ \\
\hline
\rowcolor{mygray}
\multicolumn{8}{c|}{{\textit{Medium Token Pruning} \ $\fg{(\downarrow 60\%)}$}}
& \multicolumn{8}{c}{{\textit{Medium Token Pruning} \ $\fg{(\downarrow 60\%)}$}} \\
FastV~\scriptsize{(ECCV24)}
& 49.8 & 46.4 & 41.3 & 58.2 & \textbf{51.5} & 90.0\% & $-10.0\%$
& FastV~\scriptsize{(ECCV24)} & 55.4 & 57.6 & 57.3 & 64.0 & \textbf{28.0} & 95.9\% & $-4.1\%$ \\
DART~\scriptsize{(EMNLP25)}
& 48.2 & 48.6 & 50.2 & 63.4 & 42.0 & 91.8\% & $-8.2\%$
& DART~\scriptsize{(EMNLP25)} & 52.1 & 57.4 & \textbf{59.9} & \textbf{65.7} & 27.5 & 96.7\% & $-3.3\%$ \\
Frame~\scriptsize{(2026.1)}
& 54.4 & 51.9 & \textbf{53.1} & 63.5 & 47.5 & \textbf{98.4\%} & $\textbf{-1.6\%}$
& Frame~\scriptsize{(2026.1)} & \textbf{56.5} & 59.0 & 59.0 & 65.0 & 23.0 & \textbf{96.9\%} & $\textbf{-3.1\%}$ \\
\textbf{\ours{}}
& \textbf{56.7} & \textbf{53.0} & 52.1 & \textbf{63.6} & 45.0 & \textbf{98.4\%} & $\textbf{-1.6\%}$
& \textbf{\ours{}} & 54.6 & \textbf{59.6} & 59.8 & 65.0 & 23.0 & 96.6\% & $-3.4\%$ \\
\hline
\rowcolor{mygray}
\multicolumn{8}{c|}{{\textit{Extreme Token Pruning} \ $\fg{(\downarrow 90\%)}$}}
& \multicolumn{8}{c}{{\textit{Extreme Token Pruning} \ $\fg{(\downarrow 90\%)}$}} \\
FastV~\scriptsize{(ECCV24)}
& 41.7 & \textbf{44.3} & 39.5 & 55.3 & \textbf{52.0} & 84.5\% & $-15.5\%$
& FastV~\scriptsize{(ECCV24)} & \textbf{44.7} & \textbf{53.7} & 57.0 & 58.9 & 23.0 & 88.8\% & $-11.2\%$ \\
DART~\scriptsize{(EMNLP25)}
& 41.0 & 41.5 & 53.5 & \textbf{61.7} & 45.5 & 88.5\% & $-11.5\%$
& DART~\scriptsize{(EMNLP25)} & 41.4 & 50.3 & 58.5 & 61.5 & 26.0 & 89.8\% & $-10.2\%$ \\
Frame~\scriptsize{(2026.1)}
& 42.3 & 41.0 & 54.3 & 61.1 & 48.0 & 89.6\% & $-10.4\%$
& Frame~\scriptsize{(2026.1)} & 42.4 & 49.2 & \textbf{58.8} & 62.5 & \textbf{26.5} & \textbf{90.2\%} & $\textbf{-9.8\%}$ \\
\textbf{\ours{}}
& \textbf{46.5} & \textbf{44.3} & \textbf{55.0} & 60.5 & 45.5 & \textbf{91.6\%} & $\textbf{-8.4\%}$
& \textbf{\ours{}} & 41.7 & 50.3 & \textbf{58.8} & \textbf{62.9} & 25.5 & \textbf{90.2\%} & $\textbf{-9.8\%}$ \\
\bottomrule[1.5pt]
\end{tabular}%
} 

\vspace{-1mm}
\label{tab:pruning_comparison_results_mmaupro}
\end{table*}

%% file: tables/tab_cross_model_mmaupro.tex
\begin{table}[ht]

\caption{Cross-model evaluation on MMAU-Pro with Phi-4-Multimodal. Accuracy per category. Vanilla Avg.\ set to 100\%, other Avg.\ shown as \% of Vanilla. Best Avg.\ per ratio in \textbf{bold}.}
\centering
\setlength{\tabcolsep}{4pt}
\renewcommand{\arraystretch}{0.9}
\resizebox{\columnwidth}{!}{%
\begin{tabular}{lccccc|cc}
\toprule[1.5pt]
\multirow{2}{*}{\textbf{Method}}
& \multicolumn{2}{c}{\textbf{Speech}}
& \multicolumn{3}{c|}{\textbf{Acoustic}}
& \multirow{2}{*}{\textbf{Avg.}} & \multirow{2}{*}{\textbf{Diff.}} \\
\cmidrule(lr){2-3} \cmidrule(lr){4-6}
    & $Speech$ & $VoiceChat$ & $Sound$ & $Music$ & $Spatial$ & & \\ \midrule

\rowcolor{gray!20}
\multicolumn{8}{c}{\textit{Vanilla (Phi-4-MM)}} \\
& 52.8 & 50.5 & 35.0 & 45.4 & 27.2 & 100.0\% & -- \\
\rowcolor{mygray}
\multicolumn{8}{c}{{\textit{Light Token Pruning} \ $\fg{(\downarrow 30\%)}$}} \\
FastV~\scriptsize{(ECCV24)}
& 49.8 & \textbf{50.0} & 35.9 & 45.0 & \textbf{27.2} & 99.2\% & $-0.8\%$ \\
Frame~\scriptsize{(2026.1)}
& 49.8 & \textbf{50.0} & 35.9 & 45.0 & \textbf{27.2} & 99.2\% & $-0.8\%$ \\
DART~\scriptsize{(EMNLP25)}
& 46.9 & 47.8 & \textbf{36.5} & \textbf{45.5} & 23.8 & 96.9\% & $-3.1\%$ \\
\textbf{\ours{}}
& \textbf{52.5} & 46.7 & 35.6 & \textbf{45.5} & 24.8 & \textbf{100.2\%} & $\textbf{+0.2\%}$ \\
\hline

\rowcolor{mygray}
\multicolumn{8}{c}{{\textit{Medium Token Pruning} \ $\fg{(\downarrow 60\%)}$}} \\
FastV~\scriptsize{(ECCV24)}
& 45.4 & 40.8 & 27.8 & 44.6 & \textbf{27.7} & 92.7\% & $-7.3\%$ \\
Frame~\scriptsize{(2026.1)}
& 44.9 & 41.9 & 27.4 & 43.6 & 26.7 & 92.1\% & $-7.9\%$ \\
DART~\scriptsize{(EMNLP25)}
& 48.5 & 38.0 & 34.4 & 45.0 & 22.8 & 94.2\% & $-5.8\%$ \\
\textbf{\ours{}}
& \textbf{48.8} & \textbf{45.7} & \textbf{36.0} & \textbf{45.4} & 25.7 & \textbf{97.5\%} & $\textbf{-2.5\%}$ \\

\hline
\rowcolor{mygray}
\multicolumn{8}{c}{{\textit{Extreme Token Pruning} \ $\fg{(\downarrow 90\%)}$}} \\
FastV~\scriptsize{(ECCV24)}
& 40.7 & \textbf{47.3} & 29.8 & 42.8 & 19.3 & 90.2\% & $-9.8\%$ \\
DART~\scriptsize{(EMNLP25)}
& 38.0 & 43.5 & 34.2 & \textbf{44.8} & \textbf{20.8} & 89.1\% & $-10.9\%$ \\
Frame~\scriptsize{(2026.1)}
& 38.7 & 43.5 & 27.8 & 40.2 & 14.4 & 85.8\% & $-14.2\%$ \\
\textbf{\ours{}}
& \textbf{40.9 }& 44.0 & \textbf{34.4} & 43.2 & 16.3 & \textbf{91.6\%} & $\textbf{-8.4\%}$ \\
\bottomrule[1.5pt]
\end{tabular}%
} 

\label{tab:cross_model_mmaupro}
\vspace{-4mm}
\end{table}

%% file: figures/fig_efficiency.tex
\begin{figure*}[t]
\centering
\includegraphics[width=\textwidth]{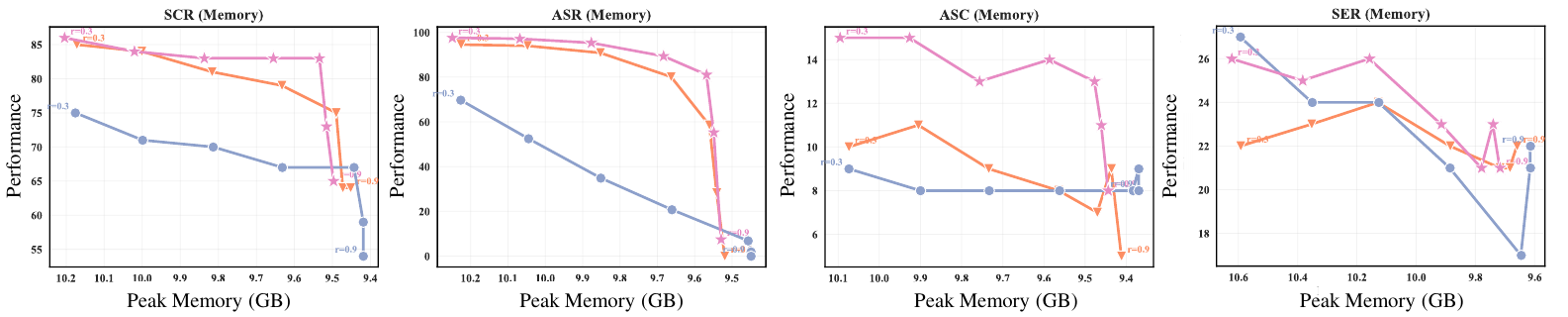}
\vspace{-6mm}
\caption{Efficiency-performance trade-offs (Qwen2.5-Omni-3B) across four AudioMarathon tasks at pruning ratios r$ \in \{0.3, 0.4, \ldots, 0.9\}$. F1-score vs.\ Peak GPU Memory (FastV omitted due to its high memory ceiling, ${\sim}15$--$20$\,GB vs.\ ${\sim}10$\,GB for all other methods). \ours{} (pink) dominates the Pareto frontier on SCR and ASR, and leads on ASC, MC, and SER.}
\vspace{-2mm}
\label{fig:efficiency}
\Description{10-panel efficiency plot showing F1-score vs latency and memory across five tasks: SCR, ASR, ASC, MC, SER.}
\end{figure*}

%% file: figures/fig_ablation_bar.tex
\begin{figure}[ht]
\centering
\includegraphics[width=\columnwidth]{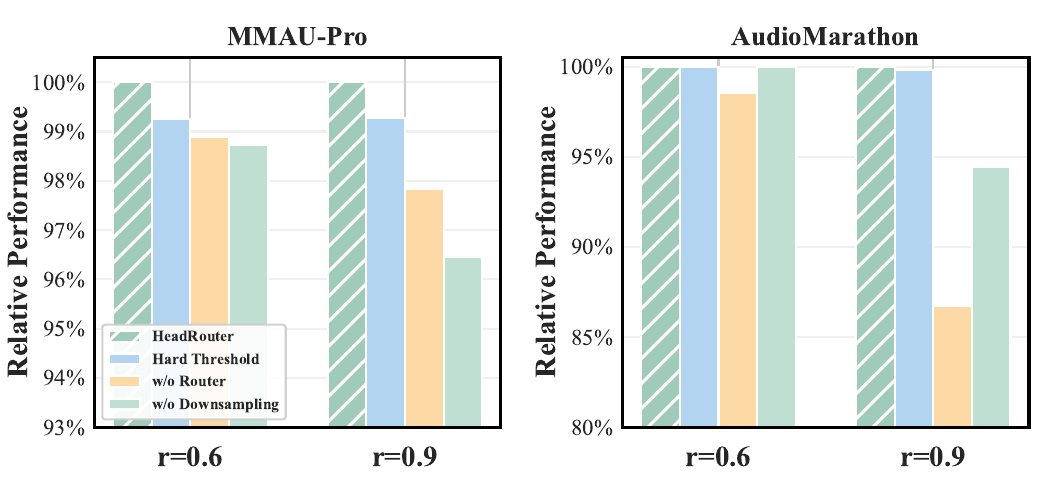}
\vspace{-4mm}
\caption{Ablation on Qwen2.5-Omni-3B at r $\in \{0.6, 0.9\}$, illustrating F1-score as a percentage of the unpruned baseline.}
\label{fig:ablation_bar}
\vspace{-3mm}
\Description{ablation results}
\end{figure}

%% file: section/7_conclusion.tex
\section{Conclusion}
\label{sec:conclusion}

We presented \textbf{\ours{}}, a training-free and input-adaptive routing mechanism for audio token pruning in large audio-language models. By revealing the distinct head-behavior patterns of semantic and acoustic tasks, and by using selectivity spread signal from QK probing as a routing signal, \ours{} dynamically mixes pre-calibrated semantic, uniform, and acoustic head-weight profiles to produce sample-specific token-importance estimates. HeadRouter matches or exceeds the vanilla baseline when removing 30\% of tokens, while all competing methods degrade, supporting that pruning may also reduce harmful and noisy tokens. These findings expose limits of importance-averaging methods and offer new insights into audio token utilization in LALMs. We will release our code publicly to support future research.